\documentclass[%
 aip, jcp,amsmath,amssymb,reprint,
]{revtex4-2}

\usepackage{graphicx}
\usepackage{dcolumn}
\usepackage{bm}
\usepackage[mathlines]{lineno}

\usepackage[utf8]{inputenc}
\usepackage[T1]{fontenc}
\usepackage{mathptmx}
\usepackage{etoolbox}

\makeatletter
\def\@email#1#2{%
 \endgroup
 \patchcmd{\titleblock@produce}
  {\frontmatter@RRAPformat}
  {\frontmatter@RRAPformat{\produce@RRAP{*#1\href{mailto:#2}{#2}}}\frontmatter@RRAPformat}
  {}{}
}%
\makeatother

\usepackage{tikz}
\usepackage{xcolor}

\begin{document}


\title{Unravelling H$_2$ chemisorption and physisorption on metal decorated graphene using quantum Monte Carlo}

\author{Yasmine S. Al-Hamdani}
\affiliation{Department of Earth Sciences, University College London, London WC1E 6BT, United Kingdom}
\affiliation{Thomas Young Centre, University College London, London WC1E 6BT, United Kingdom}
\affiliation{London Centre for Nanotechnology, University College London, London WC1E 6BT, United Kingdom}

\author{Andrea Zen}%
\affiliation{Dipartimento di Fisica Ettore Pancini, Università di Napoli Federico II, Monte S. Angelo, I-80126 Napoli, Italy}
%

%

\author{Dario Alf\`e}
\affiliation{Department of Earth Sciences, University College London, London WC1E 6BT, United Kingdom}
\affiliation{Thomas Young Centre, University College London, London WC1E 6BT, United Kingdom}
\affiliation{London Centre for Nanotechnology, University College London, London WC1E 6BT, United Kingdom}
\email{d.alfe@ucl.ac.uk}
\affiliation{Dipartimento di Fisica Ettore Pancini, Università di Napoli Federico II, Monte S. Angelo, I-80126 Napoli, Italy}

\date{\today}

\begin{abstract}
Molecular hydrogen is at the core of hydrogen energy applications and has the potential to significantly reduce the use of carbon dioxide emitting energy processes. However, hydrogen gas storage is a major bottleneck for its large-scale use as current storage methods are energy intensive. Among different storage methods, physisorbing molecular hydrogen in nanomaterials at ambient pressure and temperatures is a promising alternative -- particularly in light of the advancements in tuneable lightweight nanomaterials and high throughput screening methods. Nonetheless, understanding hydrogen adsorption in well-defined nanomaterials remains experimentally challenging and therefore, reference information is scarce despite the proliferation of works predicting hydrogen adsorption. In this work, we focus on Li, Na, Ca, and K, decorated graphene sheets as substrates for molecular hydrogen adsorption and compute the most accurate adsorption energies available to date using quantum diffusion Monte Carlo (DMC). Building on our previous insights at the density functional theory (DFT) level, we find that a weak covalent chemisorption of molecular hydrogen, known as Kubas interaction, is thermodynamically feasible on Ca decorated graphene according to DMC, in agreement with DFT. This finding is in contrast to previous DMC predictions of the 4H$_2$/Ca$^+$ gas cluster (without graphene) where chemisorption is not favoured. However, we find that the adsorption energy of hydrogen on metal decorated graphene according to a typical widely-used DFT method is not fully consistent with DMC and the discrepancies are not systematic. The reference adsorption energies reported herein can be used to find better performing work-horse methods for application in large-scale modelling of hydrogen adsorption. In addition, this work demonstrates the importance of understanding the interaction mechanisms governing adsorption as well as the corresponding absolute adsorption energies. The implications of this work affect strategies for finding suitable hydrogen storage materials and high-throughput methods.
\end{abstract}

\maketitle

\section{\label{sec:intro}Introduction}
The smallest molecule, hydrogen (H$_2$), has the potential to sharply improve sustainable energy production, by replacing fossil fuels in a variety of applications such as vehicular fuel and home heating. For example, water is the only by-product of converting H$_2$ using polymer electrolyte membrane (PEM) fuel cells and therefore its use would drastically reduce pollutants from vehicles. However, the efficient storage of hydrogen molecules is an outstanding challenge. The most used storage method currently is to pressurize H$_2$ at high ($\sim$ 700 bar) pressure inside carbon fiber tanks.\cite{Usman2022} This simple but expensive route affects the fuel economy of vehicles, detracting considerably from the benefits of hydrogen technology. Alternative storage strategies includes strong (dissociative) chemisorption, also know as the spillover effect, and 
non-dissociative physisorption, whereby H$_2$ is adsorbed on a surface as intact gas molecules. The spillover effect is typically associated with high energy barriers for hydrogen release, which also reduces the energy yield from hydrogen. Meanwhile, the spontaneous adsorption of molecular hydrogen within an energy window of $-200$ to $-400$ meV per H$_2$ molecule in a lightweight material, circumvents the need for high pressures or extreme temperatures.\cite{alhamdani2023,Li2003} Our focus is to find materials and interaction mechanisms that bring the H$_2$ adsorption energy into that range on a lightweight material. 


Graphene, carbon nanotubes, and analogous low-dimensional materials have the advantage of being lightweight with high surface areas for adsorption and being made of earth abundant elements. However, experimentally derived adsorption energies and theoretical predictions for H$_2$ on graphene and carbon nanotubes appear to be very weak, typically at less than $-50$ meV per H$_2$ molecule.\cite{Li2003,Cabria2020,al-hamdani2017cnt,Zuttel2002,Ye1999,Spyrou2013,Patchkovskii2005,Hirscher2003,Zuttel2004}, In our previous work,\cite{alhamdani2023} a density functional theory (DFT) based method predicts that metal decorating atoms on graphene, such as Li and Ca, can boost the adsorption energy by over 100 meV -- bringing it almost into the useful binding energy window for storage. However, it is well known that DFT methods suffer from the electron delocalization problem and can also be inaccurate for predicting long-range dispersion based interactions. In particular, it has previously been shown that DFT methods wrongly stabilize a covalent bond between hydrogen and group 2 metals with available 3$d$-states such as Ca, known as Kubas-type binding,\cite{Kubas2001} in the gas phase.\cite{Bajdich2010,Purwanto2011,Cha2009,Cha2011} However, it is not known if this inaccuracy carries over to the solid-state system where the metal atoms are supported by graphene. Since Kubas-type binding is predicted to be a very tunable form of chemisorption interaction in low-dimensional materials at the DFT level, we aim to establish whether it is a feasible interaction using a higher-accuracy method in this work. We use a wavefunction based method that has been shown to have benchmark accuracy, namely, quantum diffusion Monte Carlo (DMC). We predict the most promising materials indicated in our previous work with DMC and report the most accurate adsorption energies available to date for H$_2$ on pristine and metal decorated graphene. In doing so, we establish the propensity of the materials considered here to bind H$_2$. We also find that a widely-used DFT method slightly overestimates the binding for alkali metal (Li, Na, and K) decorated graphene and severely overestimates binding on Ca decorated graphene.  

\section{\label{sec:methods}Methods and Computational Setup}
\begin{figure}
\includegraphics[width=0.45\textwidth]{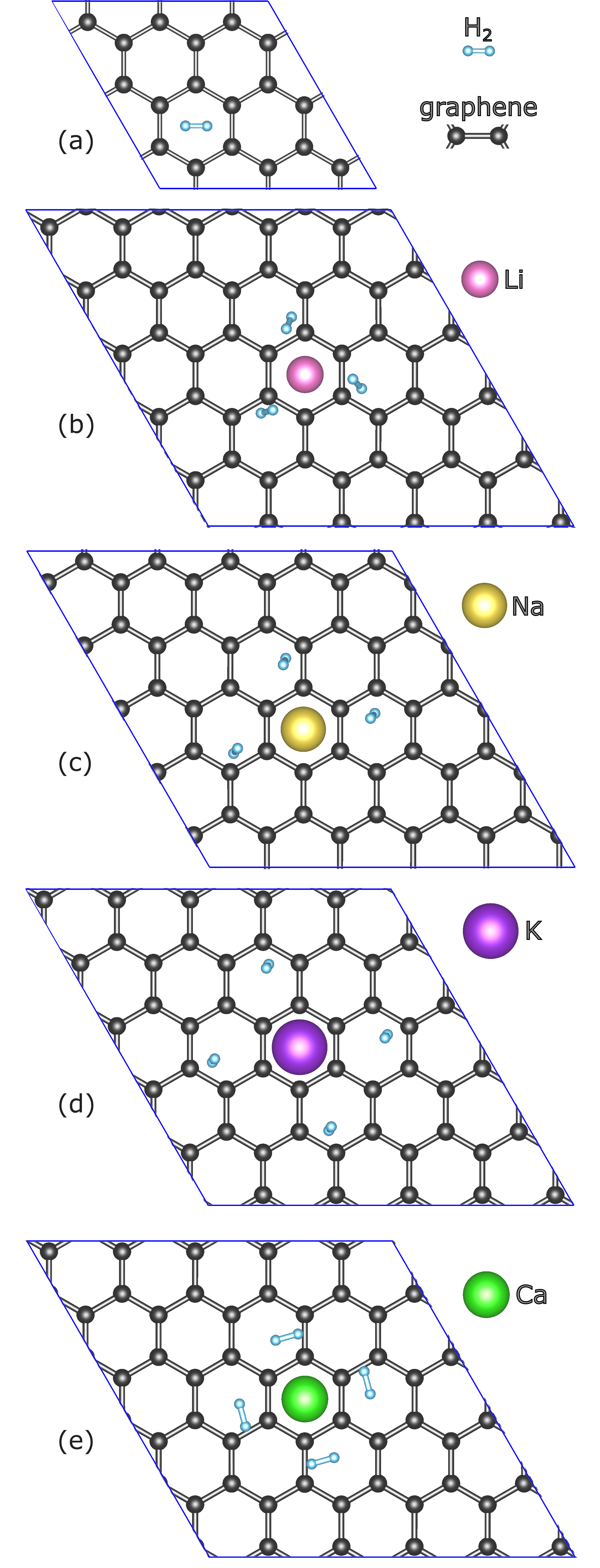}
\caption{\label{fig:fig1} The PBE+D3 geometries of (a) H$_2$ on pristine graphene (H$_2$+Gr), (b) 3H$_2$+Li@Gr, (c) 3H$_2$Na@Gr, (d) 4H$_2$+K@Gr, and (e) 4H$_2$+Ca@Gr. The unit cell is indicated by a blue bounding box.}
\end{figure}

We established PBE+D3 binding geometries for hydrogen on pristine graphene in Ref.~\onlinecite{al-hamdani2017cnt} and on Li, Na, Ca, and K decorated graphene in a recent work.\cite{alhamdani2023} PBE is a widely used generalized gradient approximation in DFT\cite{Perdew1996} and a semi-empirical two and three-body dispersion correction, D3 is added using the zero-damping function.\cite{Grimme2010} We use the optimized geometries with the most favourable binding energies per H$_2$ molecule according to PBE+D3: 3H$_2$+Li@Gr, 3H$_2$+Na@Gr, 4H$_2$+K@Gr, and 4H$_2$+Ca@Gr, as can be seen in Fig.~\ref{fig:fig1}. The adsorption energy is defined as:
\begin{equation}
    E_{ads} = (E^{tot}_{nH_2+M@Gr} - E^{tot}_{M@Gr} - nE^{tot}_{H_2}) / n
\end{equation}
where $E^{tot}_{nH_2+M@Gr}$ is the total energy of $n$ number of hydrogen molecules adsorbed on metal decorated graphene (M@Gr) where the metal can be Li, Na, K, or Ca. Correspondingly, $E^{tot}_{M@Gr}$ is the total energy of the metal decorated graphene substrate and $E^{tot}_{H_2}$ is the total energy of a gas phase hydrogen molecule. Note that the graphene sheet is a $(5\times5)$ unit cell of graphene and PAW potentials (available for the VASP 5.4.4 software package) with explicit semi-core electrons were used for the metal atoms.\cite{Kresse1993,Kresse1994,Kresse1996,Kresse1996a} In addition to the geometries previously established, we used PBE+D3 to find a physisorption minimum for 4H$_2$ molecules on Ca@Gr. This structure is also available in the Supplemental Material (SM) and the corresponding H$_2$--Ca separation distances are \textit{ca.} 3.3-3.6~\AA~ with an adsorption energy of $-81$ meV per H$_2$ molecule. In order to understand the impact of electron localization at the DFT level, we use the simplified approach to the Hubbard U method, introduced by Dudarev \textit{et al.},\cite{Dudarev1998} as implemented in VASP v.5.4.4. These heuristic calculations used U values for $3d$-states in Ca between 1 to 6 eV, which is a range that is typically applied to metal atoms. Note that we set the J value to zero in all calculations. The DFT-VASP calculations of the Ca@Gr based system undertaken in this work are computed using a $3\times3\times1$ \textbf{k}-point grid, shifted from $\Gamma$ by $(1/2, 1/2, 0)$ and 400 eV plane-wave cut-off. This yields only a 2 meV difference in the (Kubas-bound) 4H$_2$+Ca@Gr PBE+D3 adsorption energy with respect to our previous work.\cite{alhamdani2023} 

Previously, we computed the adsorption energy of a single H$_2$ molecule on pristine graphene\cite{al-hamdani2017cnt} using a ($3\times3$) unit cell of graphene and two \textbf{k}-points. Fixed-nodes were used with Slater-Jastrow type trial wavefunctions, where the orbitals for the Slater determinant were computed using PWSCF in Quantum Espresso\cite{Giannozzi2009,Giannozzi2017}. The result was $-24\pm11$ meV. Here, we follow the same computational setup but we invoke recent developments, mainly the determinant localization approximation (DLA) in QMC,\cite{Zen2019} which avoids a bias in the QMC energy from the use of different Jastrow factors in the limit of small time-steps. We used CASINO v.2.13\cite{Needs2020} (with time-steps of 0.03 and 0.05 a.u.) as well as the GPU enabled DMC algorithm in QMCPACK v.13.5.9\cite{Kim2018} (with time-steps of 0.015 and 0.05 a.u.) to obtain the hydrogen adsorption energy with stochastic errors of 2-5 meV. The results from the two QMC codes and different time-steps are all fully consistent within the stochastic error bars. See the SM for full details. 

The geometries from PBE+D3 (available in the SM) were used to perform DFT calculations with Quantum Espresso v.6.8\cite{Giannozzi2009,Giannozzi2017}. The unit cells can be seen in Fig.~\ref{fig:fig1}. More specifically, input orbitals were computed using the local density approximation (LDA) in Quantum Espresso, with ccECP pseudopotentials\cite{Wang2019,Annaberdiyev2018,Bennett2017,Bennett2018}, a 400 Ry planewave energy cut-off, and a vacuum of \textit{ca.}~15~\AA~ in the direction perpendicular to graphene. The ccECP pseudopotentials treated 3 electrons explicitly for Li, 9 electrons for Na and K, and 10 electrons for Ca. We tested the effect of using softer-core ccECP pseudopotential for Na and Li at the DFT level. We found that the PBE+D3 adsorption energy per H$_2$ molecules deviates by 5 meV for 3H$_2$+Li@Gr and 7 meV for 3H$_2$+Na@Gr due to the softer pseudopotentials. 

The orbitals obtained from DFT are used to define the nodal surface of each system in DMC and it is typically insensitive to the choice of DFT functional that is employed, however we also validate it here using two DFT functionals. LDA is a simple and efficient method that can be used for metallic systems and to validate its use here, we computed the LDA and hybrid B3LYP\cite{b3lyp} orbitals for a small gas-phase Ca$^+$+4H$_2$ cluster for comparison. The resulting DMC total energies and interaction energies based on LDA and B3LYP orbitals are statistically indistinguishable (see Appendix Section~\ref{sec:appendix-2} for further information). All calculations were spin-polarized. For the Li, Na, and K decorated graphene systems, a \textbf{k}-point mesh of $3\times3\times1$ centred on the $\Gamma$-point was used which is sufficient to converge hydrogen adsorption energies at the DFT level. Using the resulting charge density, a separate non self-consistent field (NSCF) calculation was performed at a single \textbf{k}-point to produce orbitals for the following QMC calculations. The adsorption energy based on a single \textbf{k}-point is in agreement with the adsorption energy from the full \textbf{k}-point grid for 3H$_2$+Li@Gr, 3H$_2$+Na@Gr, and 4H$_2$+K@Gr, as can be seen in the SM. However, the adsorption energy in the 4H$_2$+Ca@Gr system is more dependent on the \textbf{k}-point grid and for this system we undertook a careful analysis of the occupations at different \textbf{k}-points. We find that a  $3\times3\times1$ \textbf{k}-point grid shifted from $\Gamma$ by $(1/2, 1/2, 0)$ in reciprocal units yields a converged adsorption energy at the LDA level, with respect to a $15\times15\times1$ \textbf{k}-point mesh. Moreover, partial occupations around the Fermi energy are avoided using the shifted $3\times3\times1$ \textbf{k}-point grid which is preferential as it allows us to perform QMC calculations at separate \textbf{k}-points using only integer occupations. LDA orbitals were obtained using NSCF calculations at each \textbf{k}-point with occupations set according to the those in the LDA SCF calculation in the full \textbf{k}-point grid. Further information and justification of this protocol based on the electronic structure for Ca@Gr based systems can be found in Appendix Section~\ref{sec:appendix-1} and the SM. The planewave orbitals were localized using B-splines\cite{Alfe2004} and a meshfactor of 0.5 in QMCPACK. 

QMC calculations were performed using GPU enabled complex version of QMCPACK v.3.15.9.\cite{Kim2018}
We optimized one, two, and three-body parameters for the Jastrow factor using variational Monte Carlo (VMC) and the OneShiftOnly optimizer implemented in QMCPACK. The Jastrow factors were optimized for the adsorbed systems at $\Gamma$-point only and the element-specific parameters from these optimizations were used in the reference systems (M@Gr and H$_2$) for consistency and efficiency. Note that we also invoked the DLA scheme which removes the bias of the Jastrow factor on the energy in the limit of small time steps.\cite{Zen2019} 
We used time-steps of 0.03 a.u. for all systems in the fixed-phase DMC calculations and we tested additional time-steps of 0.06, 0.01, 0.005 a.u. See SM for a full account of DMC time-step convergence tests. In addition, we used 21,000 total walkers across 150 GPU accelerators. Workflows were partially automated using NEXUS.\cite{Krogel2016}

\section{\label{sec:res}Results}
In the following we report the most accurate hydrogen adsorption energies available to date for metal decorated graphene sheets using fixed-phase DMC and establish the existence of a Kubas-bound chemisorption minimum on Ca@Gr. We also report the adsorption energy of hydrogen on pristine graphene with half the stochastic uncertainty from previous work.\cite{al-hamdani2017cnt} In Section~\ref{sec:res-mgr}, we report the DMC adsorption energies for fixed structures found from the screening of hydrogen molecules on M@Gr systems.\cite{alhamdani2023} In Section~\ref{sec:res-phys} we uncover the difference between chemisorption and physisorption of H$_2$ on Ca@Gr with DMC  in order to understand the reliability of DFT methods for Kubas-type binding. In Section~\ref{sec:res_dftu} we provide further heuristic insights on the effect of electron delocalization on the adsorption energy of hydrogen molecules in Kubas-type bonding using the Hubbard U method in DFT.

\subsection{\label{sec:res-mgr}Reference H$_2$ adsorption energies on Li, Na, Ca, and K decorated graphene}

\begin{figure}[ht]
\includegraphics[width=0.45\textwidth]{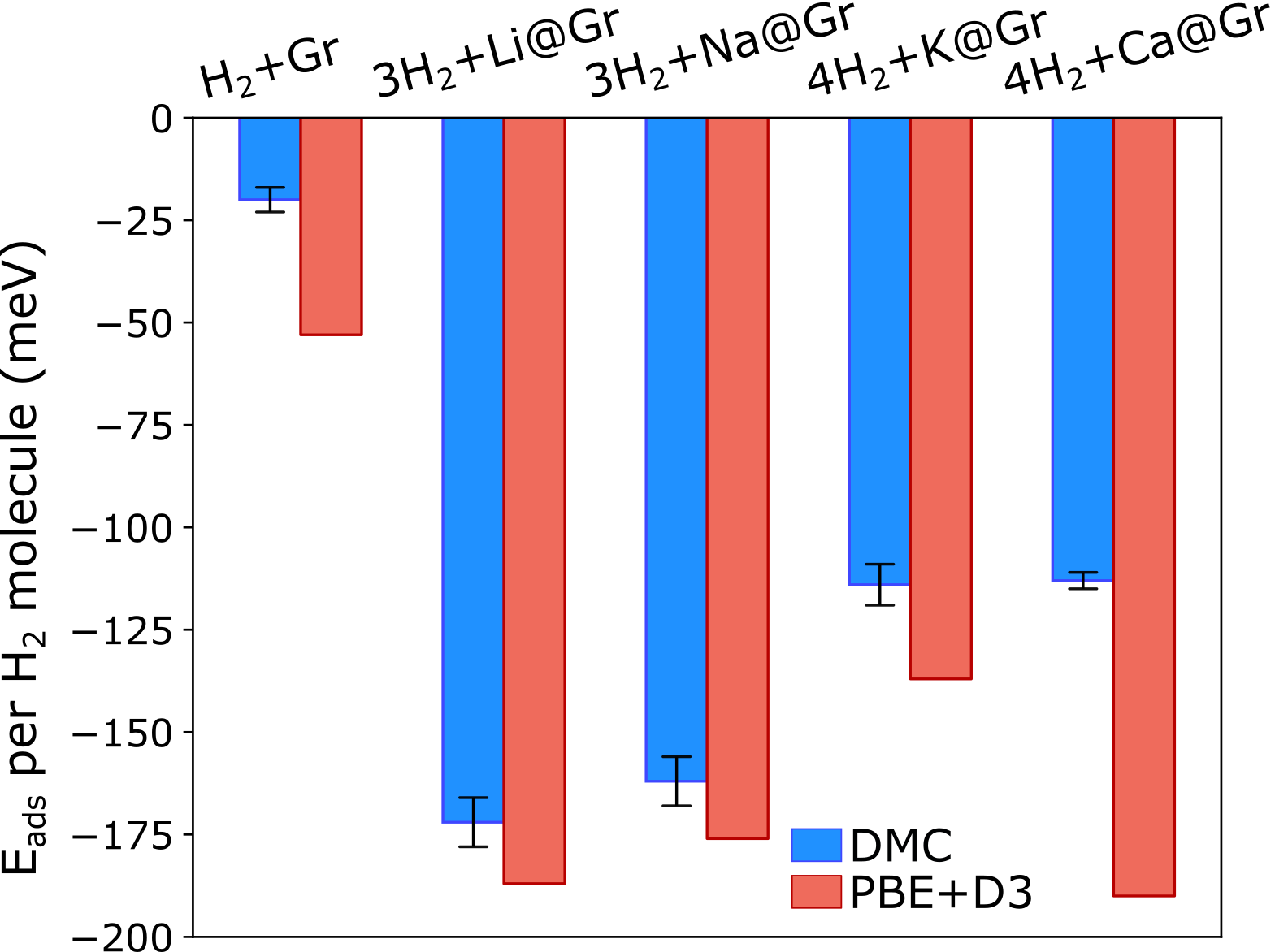}
\caption{\label{fig:res1} The DMC and PBE+D3 adsorption energy of H$_2$ on pristine graphene (Gr), Li@Gr, Na@Gr, K@Gr and Ca@Gr, in meV. We report the best available DMC result per system.}
\end{figure}

The DMC and PBE+D3 predictions of H$_2$ adsorption on pristine graphene and adatom decorated graphene are reported in Table~\ref{tab:table1} and shown in Fig.~\ref{fig:res1}. 
In addition, we report a more precise prediction of the hydrogen adsorption energy on pristine graphene ($-20\pm3$ meV), which is consistent with our previous result ($-24\pm11$ meV).\cite{al-hamdani2017cnt}
\begin{table}[ht]
\caption{Interaction energies per H$_2$ molecule (in meV) from DMC and PBE+D3 for the most binding configurations found at the PBE+D3 level. 
The best available DMC interaction energy is reported for each system. Error bars corresponding to $1\sigma$ are given.}\label{tab:table1}
\begin{ruledtabular}
\begin{tabular}{lcc}
System                       & DMC  & PBE+D3   \\ \hline
H$_2$+Gr                     & $-20\pm3$     & $-53$\footnotemark[1] \\ 
3H$_2$+Li@Gr                 & $-172\pm6$    & $-187$          \\
3H$_2$+Na@Gr                 & $-162\pm6$    & $-176$          \\
4H$_2$+K@Gr                  & $-114\pm5$    & $-137$          \\
4H$_2$+Ca@Gr                 & $-113\pm2$    & $-190$          \\
\end{tabular}
\end{ruledtabular}
\footnotetext[1]{Value computed in Ref.~\onlinecite{al-hamdani2017cnt}}
\end{table}

The DMC reference adsorption energies allow us to assess the PBE+D3 adsorption energies, keeping in mind that PBE+D3 geometries are used throughout. 
PBE+D3 overbinds the systems we report by 9 meV in 3H$_2$+Li@Gr, 8 meV in 3H$_2$+Na@Gr, 17 meV in 4H$_2$+K@Gr, and 74 meV in 4H$_2$+Ca@Gr, per H$_2$ molecule with respect to the DMC references (outside of the 1$\sigma$ stochastic error bars in DMC). Therefore, the performance of PBE+D3 across these relatively similar materials is variable, indicating that PBE+D3 and indeed similar DFT methods may not accurately rank different materials for hydrogen storage on a large-scale. 
The large discrepancy of 70 meV between PBE+D3 and DMC for 4H$_2$+Ca@Gr is particularly noteworthy. In our previous work at the DFT level, Ca@Gr was found to be the most promising material among group 1 and group 2 adatom decorated graphene sheets for hydrogen storage. This was partly due to the favourable adsorption energy predicted using PBE+D3 and also thanks to the tuneable mechanism that underpins this binding. More specifically, it has been shown that a weak Kubas-type covalent bonding can exist between hydrogen and a metal atom with partial $d$-state occupation. We showed using DFT that this form of binding can be tuned with experimentally accessible controls such as external electric fields and substrates supporting graphene. Here, we find that PBE+D3 significantly overbinds 4H$_2$+Ca@Gr and according to DMC, Ca@Gr and K@Gr are the weakest adsorbers of hydrogen among the four materials computed in this work. In order to understand whether Kubas-type binding of hydrogen is feasible at all on Ca@Gr we have to consider the relative energy difference between the Kubas-bound 4H$_2$+Ca@Gr system and a corresponding physisorption structure on Ca@Gr. This is addressed in the following section.

\subsection{\label{sec:res-phys} Is H$_2$ Kubas-bound or physisorbed on Ca@Gr?}

We have established that the PBE+D3 absolute adsorption energy for the (chemisorbed) Kubas-bound 4H$_2$+Ca@Gr is overestimated by 74 meV with respect to DMC. Furthermore, the $-113\pm2$ meV hydrogen adsorption energy predicted by DMC is far outside the favourable window of $-200$ to $-400$ meV for hydrogen storage.\cite{alhamdani2023,Li2003} However, it is nonetheless important to establish if the Kubas-type mechanism of binding predicted at the DFT level for Ca@Gr is feasible since we previously showed that such a covalent form of interaction has potential for being tuned towards more favourable hydrogen adsorption energies. To this end, our goal is to compare the thermodynamic binding preference between the Kubas configuration and a corresponding physisorption complex. We performed PBE+D3 geometry optimizations for several different starting positions of 4H$_2$ molecules centered on Ca@Gr. The initial structures included H$_2$ in upright and flat orientations relative to the graphene sheet and Ca-H$_2$ distances of $\sim3.5$~\AA. 
\begin{figure}[ht]
\includegraphics[width=0.45\textwidth]{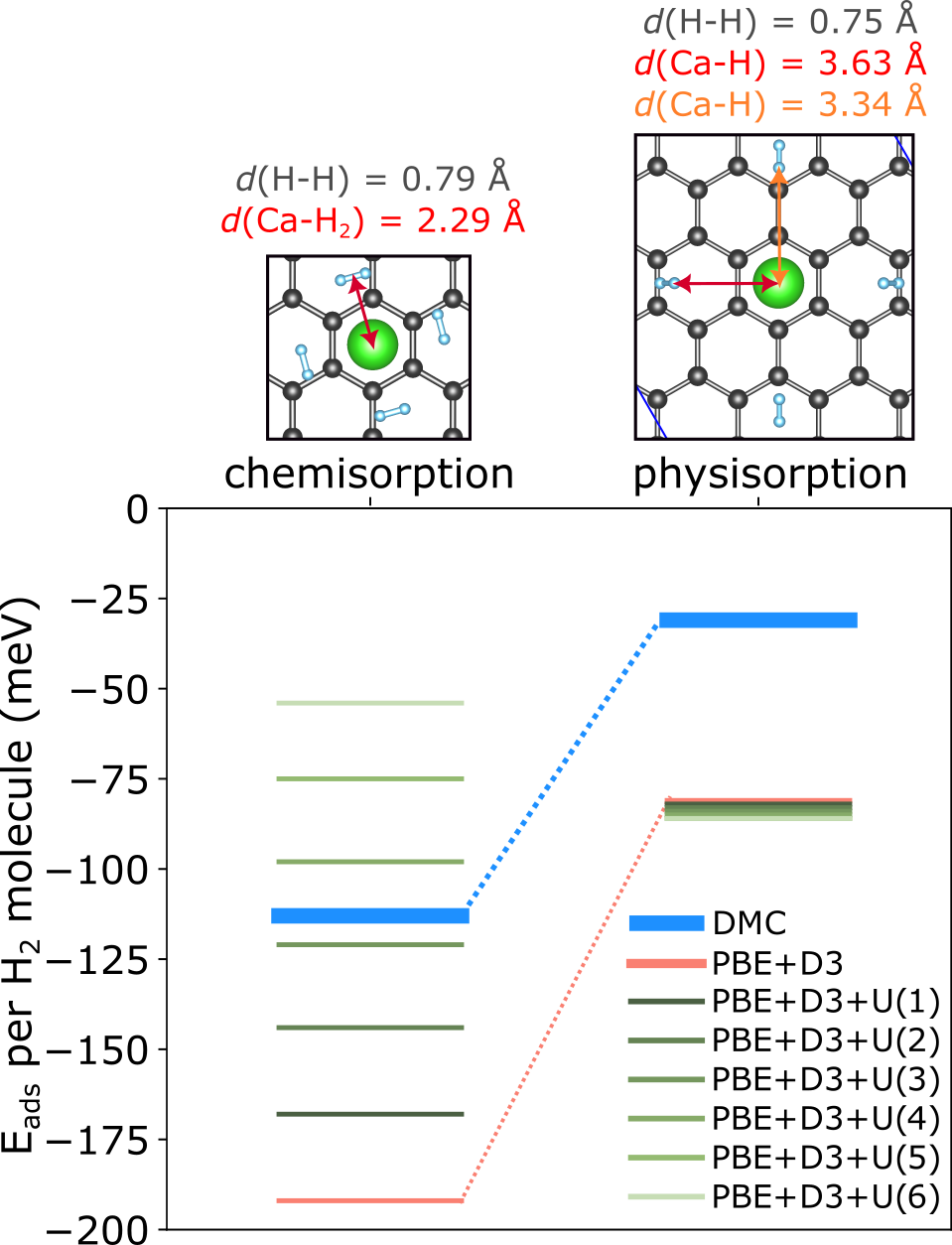}
\caption{\label{fig:res3} The PBE+D3 adsorption configurations for 4H$_2$+Ca@Gr in the Kubas chemisorption and physisorption minima and the corresponding DMC, PBE+D3, and PBE+D3+U adsorption energies, in meV per H$_2$ molecule. The U values are indicated in parenthesis. Separation distances are indicated in red and orange from the Ca adatom to the H$_2$ molecules. The H-H bond length is also reported. }
\end{figure}

The most favourable physisorption structure is shown in Fig.~\ref{fig:res3} and its adsorption energy is $-81$ meV per H$_2$ molecule with PBE+D3. 
Using the same physisorption configuration in DMC and a time-step of 0.03 a.u., the physisorption energy is $-31\pm3$ meV. 
Therefore, PBE+D3 overestimates both the chemisorption and physisorption binding energies, but their relative order is preserved in DMC indicating that Kubas-type binding is thermodynamically feasible. DMC predicts that the chemisorption state is favoured over physisorption by \textit{ca.} 80 meV at 0 K. 
As such, there is renewed potential for Kubas-type interactions to be exploited as a tuneable binding mechanism for storage of H$_2$ in metal decorated materials. 

\begin{figure}[ht]
\includegraphics[width=0.45\textwidth]{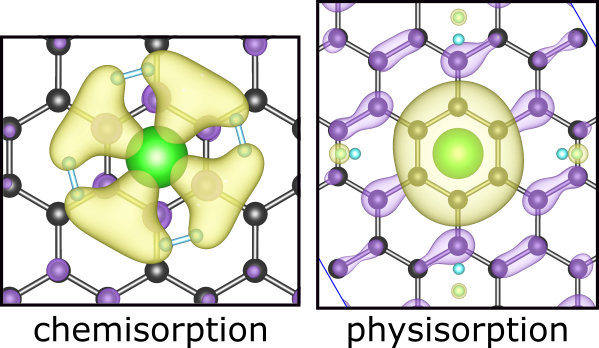}
\caption{\label{fig:res4} The 4H$_2$+Ca@Gr chemisorption and physisorption HOMOs from LDA orbitals that are used for DMC simulations. The spin-up isosurface in yellow and spin-down isosurface in purple are shown at an isosurface level of 0.005 $a{_0}^{-3/2}$. Ca is green, C atoms are grey, and H$_2$ molecules are light blue.}
\end{figure}

The considerable difference between the Kubas chemisorption and physisorption energy of 4H$_2$+Ca@Gr can be understood by noting the different electronic states that are the highest occupied molecular orbitals (HOMOs) in each configuration, as shown in Fig.~\ref{fig:res4}. Given that the Slater determinant of the DMC wavefunction is initialized by LDA orbitals (and the nodal surface remains fixed), it is evident from Fig.~\ref{fig:res4} that the chemisorption and physisorption states are starkly different in electronic configuration as well as geometry. Therefore, it is important to consider how the electronic configuration of the chemisorption state is affected by different electronic structure methods. To this end, we gauge the effect of electron localization on the electronic structure and adsorption energy of 4H$_2$+Ca@Gr in Section~\ref{sec:res_dftu}. 

\subsection{\label{sec:res_dftu}A heuristic test of electron localization using Hubbard U}
Previous DMC and coupled cluster (with single, double, and perturbative triple excitations) predictions of a Ca$^+$-4H$_2$ isolated gas cluster showed that Kubas-type binding is not thermodynamically favoured relative to physisorption and that DFT methods strongly favour Kubas-type binding incorrectly.\cite{Bajdich2010} In this work, we also computed the Ca$^+$-4H$_2$ gas cluster using our DMC protocols as well as using LDA and hybrid B3LYP. Our results are consistent with previous works and we report the findings in Appendix Section~\ref{sec:appendix-2} for the interested reader. This qualitatively wrong finding from DFT can be attributed to the delocalization error in exchange-correlation functionals, such that the stabilization from the orbital overlap of the 3$d$-state of Ca$^+$ and the H$_2$ $1\sigma^*$ orbitals is overestimated.  Importantly, this was found to be the case even using a hybrid density approximation such as B3LYP, which partially corrects for the delocalization error by including a fraction of exact exchange.  
\begin{table}[ht]
\caption{The chemisorption and physisorption minima from PBE+D3 geometry relaxations were computed with PBE+D3+U where U values from 1 to 6 eV have been used. The values in parenthesis are the adsorption energies following the full atomic relaxation with the corresponding PBE+D3+U functional. All adsorption energies are in meV per H$_2$ molecule.}\label{tab:table2}
\begin{ruledtabular}
\begin{tabular}{rcc}
U (eV) & Chemisorption & Physisorption \\
1                & $-168$ $(-168)$      & $-82$ $(-82)$     \\
2                & $-144$ $(-144)$      & $-83$ $(-83)$     \\
3                & $-121$ $(-121)$      & $-84$ $(-85)$     \\
4                & $-98 $ $(-98 )$      & $-85$ $(-87)$     \\
5                & $-75 $ $(-93 )$      & $-86$ $(-88)$     \\
6                & $-54 $ $(-97 )$      & $-86$ $(-87)$     \\ \hline
PBE+D3           & $-192$                        & $-81$                       \\
\textbf{DMC}     & $\mathbf{-113\pm2}$           & $\mathbf{-31\pm3}$        
\end{tabular}
\end{ruledtabular}
\end{table}
\begin{figure}[ht]
\includegraphics[width=0.45\textwidth]{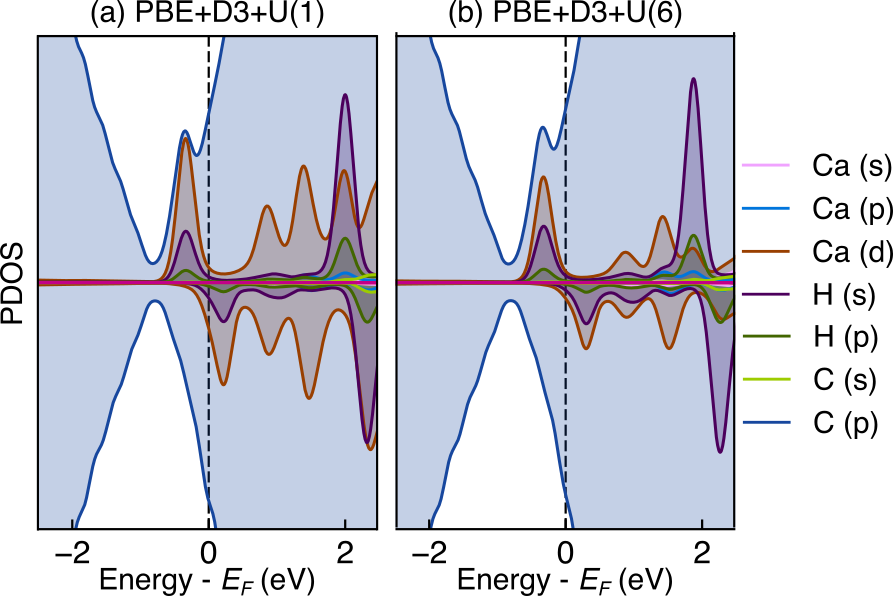}
\caption{\label{fig:res5} The PDOS of the fixed chemisorption configuration for 4H$_2$+Ca@Gr using PBE+D3+U with U value of 1 eV in (a) and 6 eV in (b). A denser \textbf{k}-point grid of $15\times15\times1$ was used and the sumo code was used for post-processing.\cite{Ganose2018}}
\end{figure}

In the materials we consider, the metal adatoms are at least partially oxidized by the graphene sheet, making the whole system metallic\cite{alhamdani2023} and thus setting them apart from the gas-cluster system computed with reference methods in the past. To understand the impact of increasing the electron localization on the chemisorption and physisorption minima of 4H$_2$+Ca@Gr, we performed heuristic PBE+D3 calculations with Hubbard U corrections for U values between 1 to 6 eV. In doing so, we find that the physisorption energy is relatively unperturbed for different U values, ranging from $-82$ to $-86$ meV, as can be seen from Fig.~\ref{fig:res3} and Table~\ref{tab:table2}. Whereas, the chemisorption minimum is strongly affected by electron localization, weakening with increasing U value, by up to $-114$ meV at U=6 eV ($-71$ meV after geometry relaxation). This can be understood by observing that the HOMO for the Kubas-bound system is dominated by the 3$d$ state of Ca overlapping with the anti-bonding 1$\sigma^*$ orbitals of H$_2$ molecules, while the physisorption HOMO is dominated by the Ca 4$s$ state, as can be seen in Fig.~\ref{fig:res4}. Importantly, the HOMO of the chemisorption state is dominated by the 3$d$ state of Ca even up to U=6 eV as can be seen in the projected density of states (PDOS) of the chemisorption system at different U values in Fig.~\ref{fig:res5}. It can also be seen that the Ca 3$d$ state is less occupied for U=6 eV than for U=1 eV. As such, the application of the U value to the $d$-state of Ca affects more strongly the binding energy of the Kubas-bound chemisorption structure. While this explains the sensitivity of the chemisorption energy to electron localization, it also suggests that different exchange-correlation functionals yield the same order of occupied electronic states and similar nodal surface for a given geometry. As fixed-phase DMC is constrained only according to the nodal surface, the implication is therefore that the DMC chemisorption energy is unaffected by different input DFT orbitals -- as is also shown in the Ca$^+$+4H$_2$ gas cluster. 

Separately, we also fully relaxed the geometries at each U value and, as can be seen from the adsorption energies in parenthesis in Table~\ref{tab:table2}, we found that the structures do not change noticeably with the exception of PBE+D3+U where U is 5 and 6 eV.
Starting from the Kubas-bound structure with high values of U results in re-configuration to a physisorption state and the bond length of H$_2$ molecules shorten. The H--H bond length at U of 5 and 6 eV is 0.76~\AA, and the distance to the Ca adatom is $\sim2.56$~\AA~ and $\sim3.04$~\AA, respectively.  In all other cases, the minimal change in structure and adsorption energy upon relaxation with different U values also suggests the use of fixed PBE+D3 geometries in DMC does not bias the results. Overall, this heuristic demonstration on the impact of localizing the electron density indicates that the Kubas-type interaction energy is sensitive to the DFT approximation used and thus suggests that such interactions are challenging to accurately predict at the DFT level. 

\section{\label{sec:conc}Conclusion}
We predicted the most accurate molecular hydrogen adsorption energies available to date on pristine graphene and metal adatom decorated graphene sheets using DMC. The DMC adsorption energy of H$_2$ on pristine graphene is $-20\pm3$ meV. This result is consistent with what was predicted with DMC in the past,\cite{al-hamdani2017cnt} but thanks to algorithmic developments and better computational efficiency, we achieved half the stochastic error here. Going beyond pristine graphene, we show that among Li, Na, K, and Ca adatoms on graphene, Li facilitates the strongest binding of molecular hydrogen, with an adsorption energy of $-172\pm6$ meV per H$_2$ molecule. Furthermore, reference DMC predicts that Ca@Gr is the weakest adsorber of H$_2$ in this subset of materials -- in stark contrast to previous DFT predictions. The broader implication of this is that a widely-used DFT method such as PBE+D3 cannot accurately rank hydrogen adsorption within a small subset of materials with different adatoms and therefore cannot be expected to deliver very accurate predictions in large-scale materials screening. However, we additionally computed a physisorption minimum for 4H$_2$ on Ca@Gr with PBE+D3 and DMC and ascertained that a chemisorption minimum is thermodynamically favoured. We have previously shown that the chemisorption minimum is underpinned by Kubas-type covalent binding and is therefore modifiable by external controls such as electric fields and substrate materials supporting the graphene sheet. As a result, the confirmation from DMC that this form of binding is thermodynamically feasible provides support for further work on exploiting Kubas-type interactions to boost the hydrogen adsorption energy for hydrogen storage applications. 
    
\appendix 
\section{Sampling the Brillouin zone in DMC using information from DFT}\label{sec:appendix-1}
The \textbf{k}-point convergence of the adsorption energy on Li, Na, and K decorated graphene is achieved with a single \textbf{k}-point with respect to a fully converged $\Gamma$ centred $5\times5\times1$ \textbf{k}-point grid, as shown in Table~\ref{atab:table3}. The single \textbf{k}-point calculations were performed non-self consistently using the fully converged charge densities with LDA and VASP v5.4.4. Although the adsorption energy cannot be obtained directly from a non-self consistent calculation using Quantum Espresso, we found that the adsorption energies are in agreement between the two codes at these \textbf{k}-point grids and that the same convergence is achieved. Using this information, we chose the K point to produce orbitals for the DMC adsorption energy computations of 3H$_2$+Li@Gr, 3H$_2$+Na@Gr, and 4H$_2$+K@Gr.
\begin{table}[ht]
\begin{ruledtabular}
\caption{Adsorption energies per H$_2$ molecule (in meV) from LDA using fully converged $3\times3\times1$ and $5\times5\times1$ \textbf{k}-point grids and non-self consistently (using the converged charge density) at the $\Gamma$ and K points.}\label{atab:table3}
\begin{tabular}{lrrr}
 & 3H$_2$+Li@Gr & 3H$_2$+Na@Gr & 4H$_2$+K@Gr  \\ \hline
$\Gamma$          & -232 & -212 & -168    \\
K                 & -235 & -215 & -172    \\
$3\times3\times1$ & -234 & -215 & -172     \\
$5\times5\times1$ & -234 & -215 & -172 
\end{tabular}
\end{ruledtabular}
\end{table}

The 4H$_2$+Ca@Gr system has a more complex convergence with respect to \textbf{k}-point sampling and we found that the adsorption energy at each (non-self consistently computed) \textbf{k}-point varies significantly from the fully converged adsorption energy. This necessitates the twist-averaging of DMC energies with orbitals obtained at different \textbf{k}-points. In order to find the most accurate (and feasible) grid for twist-averaging we analysed the results of several \textbf{k}-point grids and the electron occupancies across a dense $15\times15\times1$ grid for the three configurations: Kubas-bound 4H$_2$+Ca@Gr, physisorbed 4H$_2$+Ca@Gr, and the Ca@Gr substrate. The electron occupancy at different points in reciprocal space in the first Brillouin zone can be seen in Fig.~\ref{afig:bz}. In DFT, non-integer electron occupations are possible thanks to smearing across a set of \textbf{k}-points, however, in our real-space QMC simulations only integer electron occupations are allowed. As a result, the most suitable \textbf{k}-point grids at the DFT level correspond to those that avoid non-integer occupations and achieve good convergence for the adsorption energy. The $3\times3\times1$ grid that is shifted by $(1/2,1/2,0)$ from the $\Gamma$ point centre is well-converged for the adsorption energy and has mostly integer occupations at each \textbf{k}-point, as can be seen from Table~\ref{atab:table4} and Fig.~\ref{afig:bz}. Out of the 9 \textbf{k}-points, there are only two \textbf{k}-points with 1/2 electron each at the LDA level, and this is the case only for the physisorbed 4H$_2$+Ca@Gr and Ca@Gr substrate systems. The two \textbf{k}-points with 1/2 electron occupation are equivalent in energy and as such, we can assign one full electron to only one of the \textbf{k}-points and $1/2$ an electron less on the other \textbf{k}-point. In doing so, the correct total number of electrons and total energy is maintained when averaged. We report the occupancies from LDA and those used at the QMC level in Table~\ref{atab:table4}.

\onecolumngrid
\begin{center}
\begin{table}[ht]
\begin{ruledtabular}
\caption{The adsorption energy and electron population from the $3\times3\times1$ off-centred grid and at the corresponding separate \textbf{k}-points. The adsorption energy at each individual \textbf{k}-point is computed using NSCF calculations based on the converged charge density and the corresponding integer number of electrons in each row. The adsorption energy is reported for the Kubas-bound chemisorbed 4H$_2$+Ca@Gr system. }\label{atab:table4}
\begin{tabular}{lcccccc}
\textbf{k-points} &
  \multicolumn{1}{c}{E$_{ads}$ (meV)} &
  \multicolumn{1}{c}{Chem. 4H$_2$+Ca@Gr} &
  \multicolumn{2}{c}{Phys. 4H$_2$+Ca@Gr} &
  \multicolumn{2}{c}{Ca@Gr} \\
SCF@ $3\times3\times1$ & -363 &
  \multicolumn{1}{l}{LDA (QE) and QMC} &
  \multicolumn{1}{c}{LDA (QE)} &
  \multicolumn{1}{c}{QMC} &
  \multicolumn{1}{c}{LDA (QE)} &
  \multicolumn{1}{c}{QMC} \\ \hline
1/6, 1/6, 0   & -935 & 218 & 217.5 & 217 & 209.5 & 209 \\
1/2, 1/6, 0   & 176  & 218 & 219   & 219 & 211.0 & 211 \\
-1/6, 1/6, 0  & -882 & 218 & 217   & 217 & 209.0 & 209 \\
1/6, 1/2, 0   & -262 & 218 & 219   & 219 & 211.0 & 211 \\
1/2, 1/2, 0   & -423 & 218 & 217   & 217 & 209.0 & 209 \\
-1/6, 1/2, 0  & 177  & 218 & 219   & 219 & 211.0 & 211 \\
1/6, -1/6, 0  & -882 & 218 & 217   & 217 & 209.0 & 209 \\
1/2, -1/6, 0  & 176  & 218 & 219   & 219 & 211.0 & 211 \\
-1/6, -1/6, 0 & -412 & 218 & 217.5 & 218 & 209.5 & 210 \\ \hline
Average       & -363 & 218 & 218   & 218 & 210   & 210
\end{tabular}
\end{ruledtabular}
\end{table}
\begin{figure}
\includegraphics[width=0.9\textwidth]{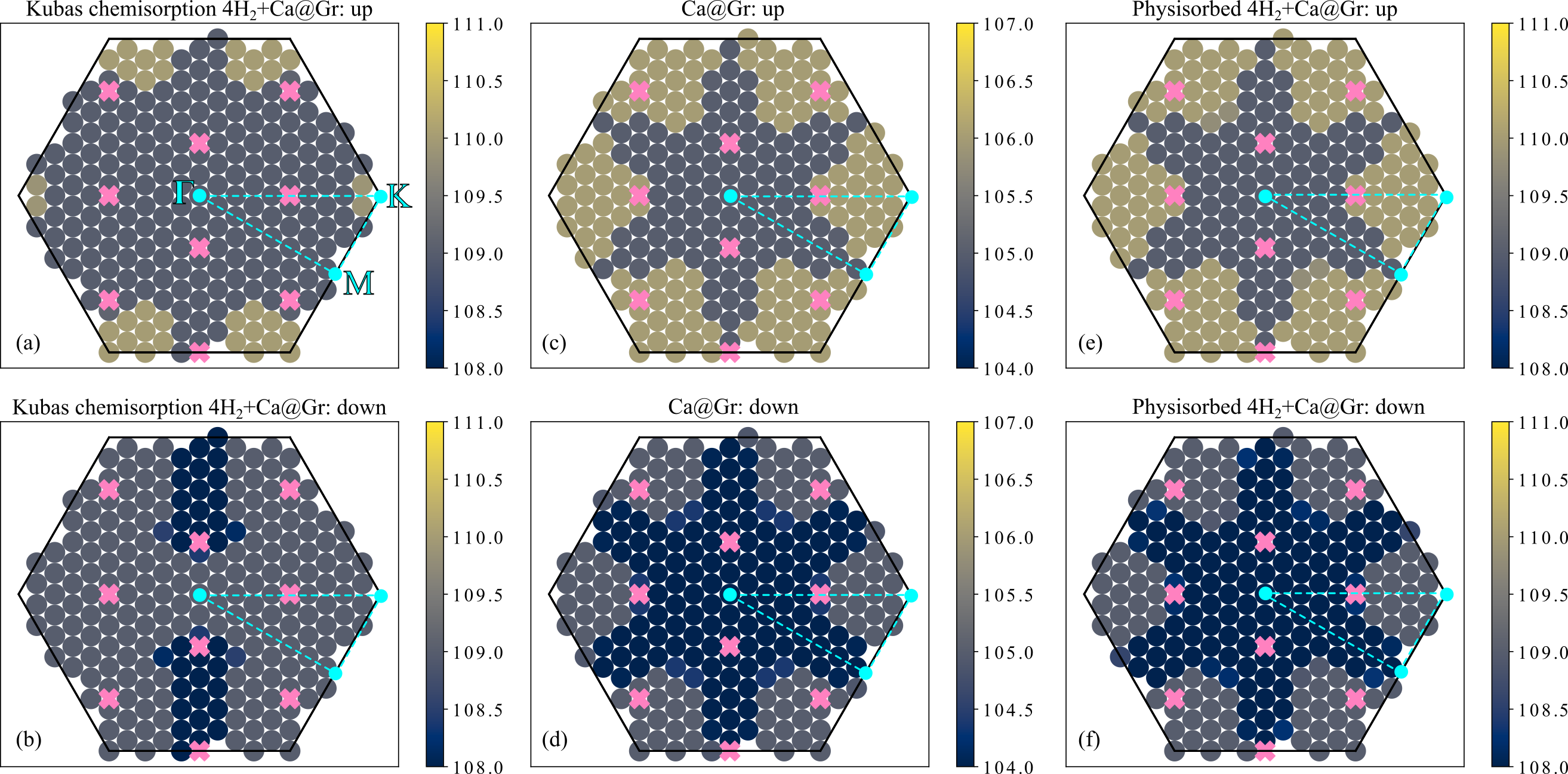}
\caption{\label{afig:bz} The occupations at each \textbf{k}-point in the first Brillouin zone according to the LDA with a $15\times15\times1$ \textbf{k}-point mesh centred on the $\Gamma$ point for the Kubas-bound chemisorbed 4H$_2$+Ca@Gr (a,b), Ca@Gr (c,d), and physisorbed 4H$_2$+Ca@Gr (e,f) configurations. Upper panels show spin-up electrons and the lower panel shows spin-down electrons. The $\Gamma$, M, and K points are indicated in light blue. The pink circles mark the $3\times3\times1$ grid shifted by $(1/2, 1/2, 0)$ from the $\Gamma$ point, which is the grid used in LDA and DMC calculations to compute the adsorption energy of 4H$_2$ on Ca@Gr.}
\end{figure}
\end{center}
\twocolumngrid

\section{The 4H$_2$+Ca$^+$ gas cluster}\label{sec:appendix-2}

\begin{figure}
\includegraphics[width=0.45\textwidth]{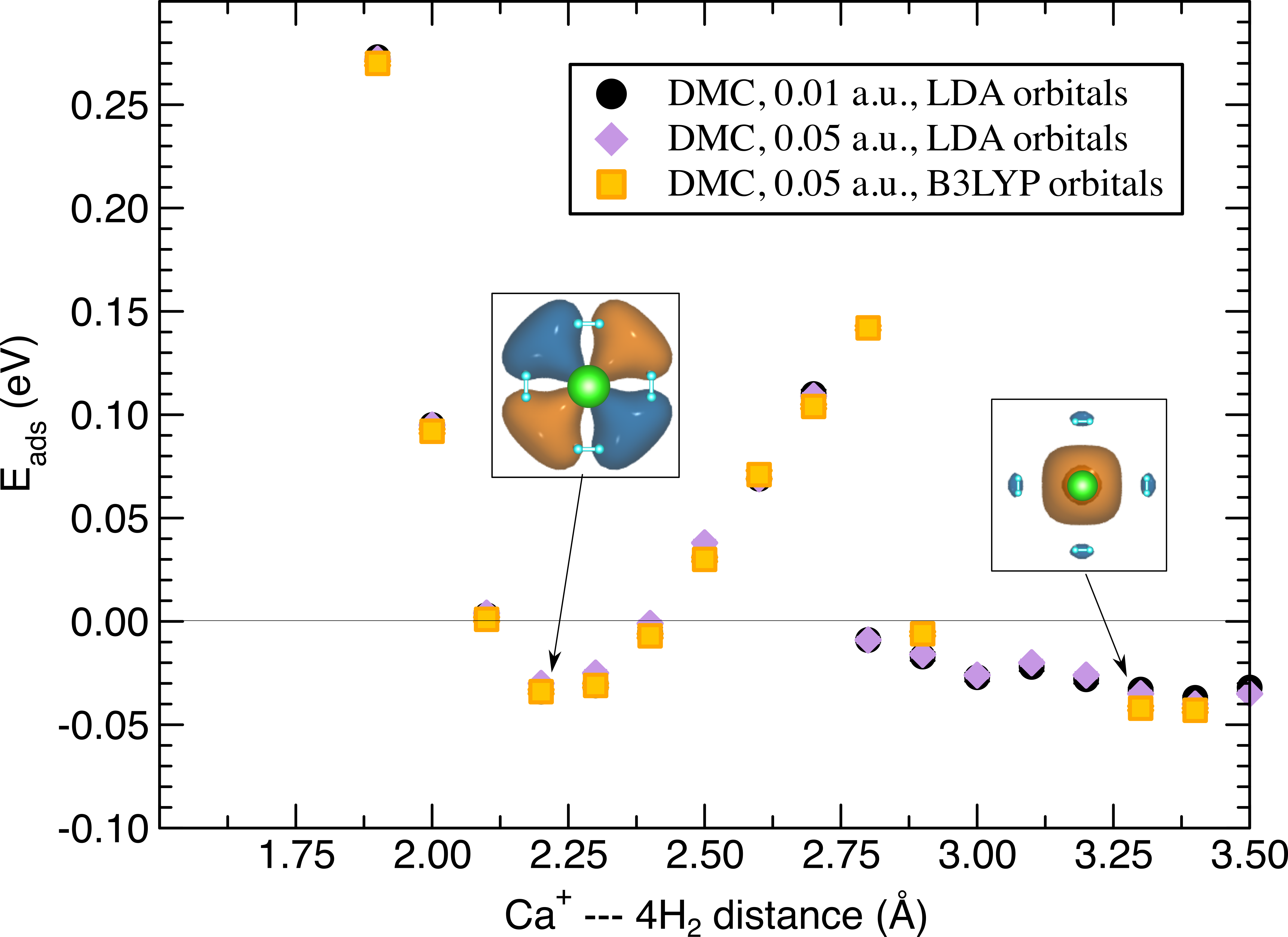}
\caption{\label{afig:gascluster} Interaction energy curves from DMC with LDA and B3LYP orbitals for the Ca$^+$---4H$_2$ gas cluster. DMC using LDA orbitals with 0.01 and 0.05 a.u. time-steps are shown. The stochastic uncertainty at each point is smaller than the marker. The insets show the B3LYP HOMO at 2.2 (d(H--H)=0.77~\AA) and 3.3~\AA~(d(H--H)=0.75~\AA) separation distance from the Ca$^+$ cation, respectively.}
\end{figure}

Bajdich \textit{et al.} have previously computed the interaction energy curve for a 4H$_2$+Ca$^+$ gas cluster, where the 3d-state of Ca is also found to be partially occupied with some methods, enabling a Kubas-type binding.\cite{Bajdich2010} We have verified this using ccECP pseudopotentials and the QMCPACK code in this work, see Fig.~\ref{afig:gascluster}. The gas cluster geometry is obtained from a constrained B3LYP geometry optimization, using ORCA and def2[3/4] basis set extrapolations. The constraint was to keep the square planar orientation of the complex with Ca at the centre, while the H--H bond length is allowed to relax. In the work of Bajdich \textit{et al.}, the H--H bond length was kept fixed for different Ca-4H$_2$ separation distances instead. However, Purwanto \textit{et al.} verified that the impact of flexible H--H bonds is small while predicting the gas cluster with auxilliary field QMC.\cite{Purwanto2011}

We used Quantum Espresso to produce LDA and B3LYP orbitals for DMC with 600 Ry planewave cut-off and the gas cluster in a unit cell box ($15\times15\times15$)~\AA. We find that the B3LYP and LDA orbitals yield near-indistinguishable interaction energies across the separation distances where physisorption and Kubas-type binding may occur. In addition, the DMC total energy with LDA and B3LYP orbitals is statistically indistinguishable, e.g. at a separation distance of 2.2~\AA~ the DMC total energy with LDA orbitals is $-1119.412\pm4$ eV while it is $-1119.408\pm3$ with B3LYP. Moreover, we confirm that DMC does not clearly favour Kubas-type binding in the gas cluster and this is independent of the exchange-correlation functional used to obtain orbitals. The main effect of the underlying orbitals (B3LYP and LDA) is to shift the distance at which the HOMO switches from the 3d orbital of Ca (corresponding to the first minimum) to the 4s orbital (corresponding to the second minimum), as can be seen from Fig.~\ref{afig:gascluster}. More specifically, the DMC-LDA HOMO at 2.8~\AA~ is mostly Ca-3$d$ while DMC-B3LYP HOMO at this distance has Ca-4$s$ character. However, near the interaction energy minima, LDA and B3LYP orbitals are consistent and therefore the DMC energies are in good agreement. Our calculations are therefore consistent with previous QMC works.\cite{Bajdich2010,Purwanto2011} 

\begin{acknowledgments}
Y.S.A. is supported by Leverhulme grant no. RPG-2020-038. A.Z. also acknowledges support by RPG-2020-038. The authors acknowledge the use of the UCL Kathleen High Performance Computing Facility (Kathleen@UCL), and associated support services, in the completion of this work. 
This research used resources of the Oak Ridge Leadership Computing Facility at the Oak Ridge National Laboratory, which is supported by the Office of Science of the U.S. Department of Energy under Contract No. DE-AC05-00OR22725). Calculations were also performed using the Cambridge Service for Data Driven Discovery (CSD3) operated by the University of Cambridge Research Computing Service (www.csd3.cam.ac.uk), provided by Dell EMC and Intel using Tier-2 funding from the Engineering and Physical Sciences Research Council (capital grant EP/T022159/1 and EP/P020259/1). This work also used the ARCHER UK National Supercomputing Service (https://www.archer2.ac.uk), the United Kingdom Car Parrinello (UKCP) consortium (EP/ F036884/1). Finally, the authors would like to thank Dr. J. Kane Shenton for help in developing the Python visualisation scripts used in this work. 
\end{acknowledgments}

\section*{Data Availability Statement}


\begin{center}
\renewcommand\arraystretch{1.2}
\begin{tabular}{| >{\raggedright\arraybackslash}p{0.3\linewidth} | >{\raggedright\arraybackslash}p{0.65\linewidth} |}
\hline
\textbf{AVAILABILITY OF DATA} & \textbf{STATEMENT OF DATA AVAILABILITY}\\  
\hline
Data available on request from the authors
&
The data that support the findings of this study are available from the corresponding author upon reasonable request.
\\\hline
\end{tabular}
\end{center}

\appendix

\bibliography{bibliography}

\end{document}


\title{Supplemental Material for:\\
Unravelling H$_2$ chemisorption and physisorption on metal decorated graphene using quantum Monte Carlo}

\author{Yasmine S. Al-Hamdani}
\affiliation{Department of Earth Sciences, University College London, London WC1E 6BT, United Kingdom}
\affiliation{Thomas Young Centre, University College London, London WC1E 6BT, United Kingdom}
\affiliation{London Centre for Nanotechnology, University College London, London WC1E 6BT, United Kingdom}
\author{Andrea Zen}%
\affiliation{Dipartimento di Fisica Ettore Pancini, Università di Napoli Federico II, Monte S. Angelo, I-80126 Napoli, Italy}
%
%
\author{Dario Alf\`e}
\affiliation{Dipartimento di Fisica Ettore Pancini, Università di Napoli Federico II, Monte S. Angelo, I-80126 Napoli, Italy}
\affiliation{Department of Earth Sciences, University College London, London WC1E 6BT, United Kingdom}
\affiliation{Thomas Young Centre, University College London, London WC1E 6BT, United Kingdom}
\affiliation{London Centre for Nanotechnology, University College London, London WC1E 6BT, United Kingdom}

\maketitle

\tableofcontents

\setcounter{section}{0}
\renewcommand{\thesection}{S\arabic{section}}%
\setcounter{table}{0}
\renewcommand{\thetable}{S\arabic{table}}%
\setcounter{figure}{0}
\renewcommand{\thefigure}{S\arabic{figure}}%

\def\bibsection{\section*{Supplementary References}}

\date{\today}


\section*{Summary}
The supplemental material herein is intended to support ``Reference quantum Monte Carlo adsorption energies for H$_2$ molecules on metal decorated graphene'' and provide additional details that may interest some readers. We provide details on time-step convergence in DMC, estimate of two-body finite size effects in QMC, and the band structure of the Ca decorated graphene systems.

\newpage
\section{Time-step convergence in DMC}
The H$_2$+Gr system was previously computed using the CASINO code with Trail and Needs pseudopotentials\cite{Trail2005,Trail2005b} and a time-step of 0.015 a.u. in DMC, based on LDA orbitals. In Table~\ref{stab:table1} we report additional calculations from CASINO and QMCPACK at different time-steps, demonstrating agreement between codes and time-steps. 
\begin{table}[ht]
\caption{Interaction energy of a H$_2$ molecule on graphene (in meV) from DMC at different time-steps using CASINO v2.13 and GPU enabled DMC algorithm in QMCPACK v.13.5.9. The structure is from Ref.~\cite{al-hamdani2017cnt}. Error bars corresponding to $1\sigma$ are given.}\label{stab:table1}
\begin{ruledtabular}
\begin{tabular}{lcccc}
System          & {0.015 a.u.}                 & {0.03 a.u.}  & {0.05 a.u.}   \\ \hline
CASINO (CPU)    & $-24\pm11$\footnotemark[1]   & $-20\pm3$    & $-21\pm5$     \\ 
QMCPACK (GPU)   & $-20\pm3$                    & n/a          & $-19\pm2$     \\
\end{tabular}
\end{ruledtabular}
\footnotetext[1]{Value computed in Ref.~\onlinecite{al-hamdani2017cnt} without DLA and with different pseudopotentials.}
\end{table}

It can be seen from Table~\ref{stab:table2} that there is only a small time-step dependence of the DMC adsorption energies using 0.03 and 0.06 a.u. time-steps. We computed the chemisorbed 4H$_2$+Ca@Gr system using 0.005 a.u. time-step, resulting in an adsorption energy of $-111\pm5$ meV, which is within the 1$\sigma$ stochastic error of the result from 0.03 a.u. 

\begin{table}[ht]
\caption{Interaction energies per H$_2$ molecule (in meV) from DMC at different time-steps for the most binding configurations found at the PBE+D3 level on Li, Na, K, and Ca decorated graphene. 
Error bars corresponding to $1\sigma$ are given.}\label{stab:table2}
\begin{ruledtabular}
\begin{tabular}{lcccc}
System                & {0.005 a.u.}    & {0.01 a.u.}     & {0.03 a.u.}  & {0.06 a.u.}     \\ \hline
3H$_2$+Li@Gr          & n/a             & n/a             & $-178\pm7$     & $-176\pm3$                 \\
3H$_2$+Na@Gr          & $-156\pm13$           & $-147\pm9$      & $-162\pm6$    & n/a                 \\
4H$_2$+K@Gr           & n/a             & n/a             & $-114\pm5$     &    n/a                     \\
4H$_2$+Ca@Gr          & $-111\pm5$      & $-103\pm5$      & $-113\pm2$     &    n/a                     \\
\end{tabular}
\end{ruledtabular}
\end{table}


\section{Estimate of two-body finite size effects in QMC from DFT}
Two-body finite size effects can affect QMC energies and there are a few of methods that can be used to estimate this contribution, \textit{e.g.} the Model Periodic Coulomb (MPC) potential and Chiesa correction\cite{Williamson1997,Chiesa2006} in QMC and the Kwee-Zhang-Krakauer (KZK) functional\cite{Kwee2008} in DFT, which are implemented for non-spin polarized systems in QMCPACK and Quantum Espresso, respectively. 
Our production calculations are spin-polarized at the DFT and QMC level to properly represent the electronic structures that result from metal adatoms adsorbing on graphene. As such, we estimate two-body finite effects using the KZK method in a non-spin polarized calculation of the 4H$_2$+Ca@Gr adsorption energy. We find the difference between KZK and LDA to be 3 meV in the adsorption energy per H$_2$ molecule. We consider this to be near-negligible and within the stochastic error that we obtain for QMC adsorption energies. Note also that the difference between spin polarized and non-spin polarized adsorption energy for this system is \textit{ca.} 30 meV at the LDA level and therefore it is important to perform spin-polarized simulations.

\section{DFT band structure of Ca decorated graphene systems}
\begin{figure}[ht]
\includegraphics[width=0.9\textwidth]{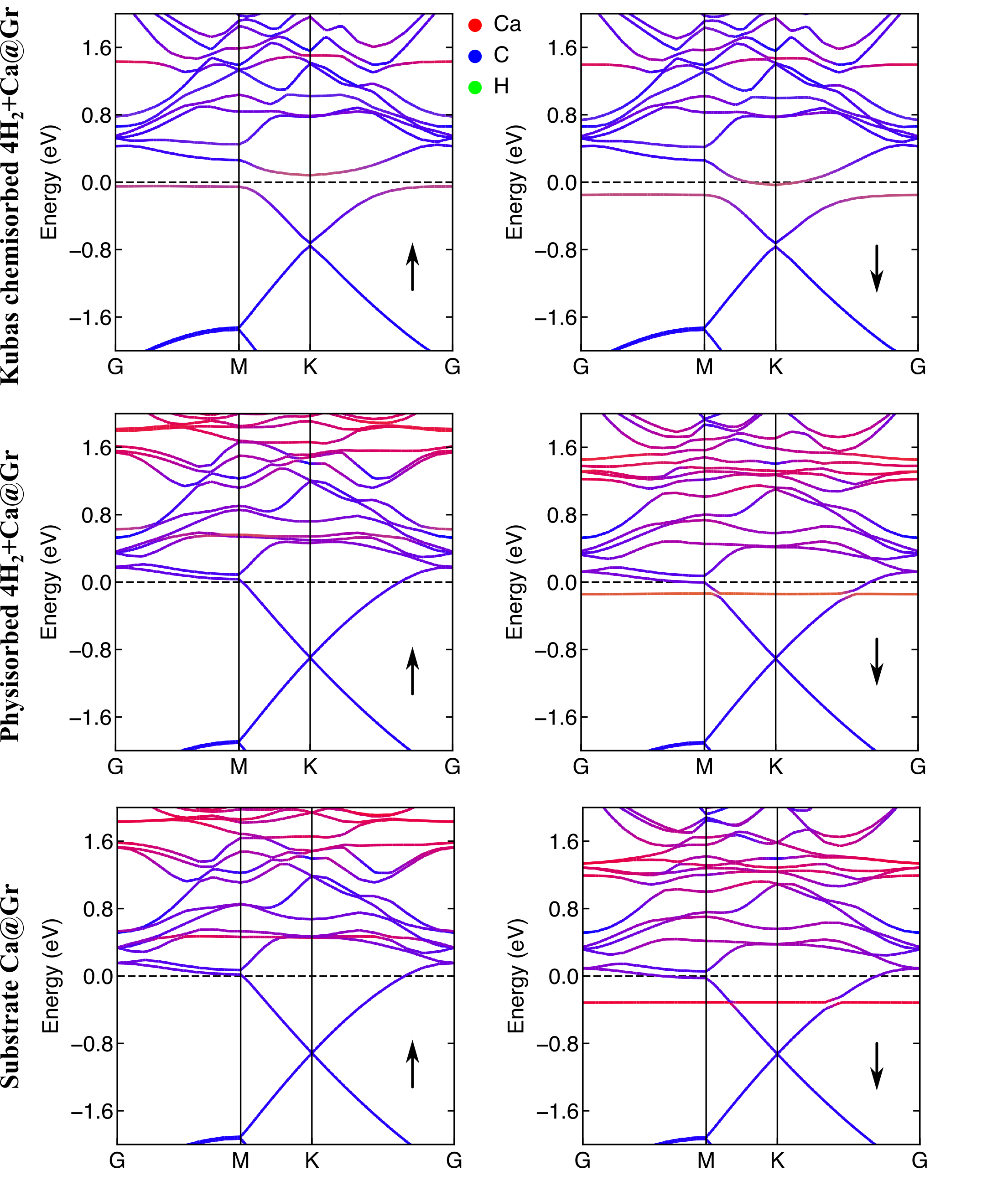}
\caption{\label{sfig:band_structure} The band structure along the $\Gamma$-K-M \textbf{k}-point path according to the LDA with the charge density converged using a $15\times15\times1$ \textbf{k}-point mesh centred on the $\Gamma$ point. The top panel shows band structure of the Kubas-bound chemisorbed 4H$_2$+Ca@Gr system, the middle panel shows physisorbed 4H$_2$+Ca@Gr, and the bottom panel shows the substrate, Ca@Gr. Spin-up band structures are shown on the left and spin-down electron band structures are shown on the right.}
\end{figure}

The band structure has been computed non-self consistently for the $\Gamma$-M-K path using the charge density from a $15\times15\times1$ \textbf{k}-point mesh (computed self-consistently). See Fig~.\ref{sfig:band_structure} for the spin up and spin down band structures of chemisorbed 4H$_2$+Ca@Gr, physisorbed 4H$_2$+Ca@Gr, and Ca@Gr. It can be seen that the band structures of physisorbed 4H$_2$+Ca@Gr and Ca@Gr are qualitatively alike due to the weak perturbation on the electronic structure from the weak interaction with H$_2$ molecules. On the other hand, the band structure of chemisorbed 4H$_2$+Ca@Gr is markedly different due to the Ca (d) state around the Fermi energy.

\pagebreak

\bibliography{bibliography}